\documentclass[aps,prl,showpacs,twocolumn,preprintnumbers,groupedaddress]{revtex4-1}
\usepackage{graphicx,amsmath,amssymb}
\usepackage{bm} 
\usepackage{mathpazo,times}
\usepackage[colorlinks=true,citecolor=blue]{hyperref}

\newcommand\partialderiv[3][]{\frac{\partial^{#1}#2}{\partial {#3}^{#1}}}
\def\Real{\mathbb{R}}

\def\Im{\mathrm{Im}}
\def\DDy#1{\frac{D#1}{Dy}}
\def\ddy#1{\partialderiv{#1\!}{y}\,}
\let\@ = \mathbf
\def\d{\mathrm{d}}
\def\e{\mathrm{e}}
\def\diag{\mathop{\rm diag}\nolimits}
\let\~=\tilde
\makeatother

\def\[{\begin{equation}}
\def\]{\end{equation}}
\def\be{\begin{equation}}
\def\ee{\end{equation}}
\def\bse{\begin{subequations}}
\def\ese{\end{subequations}}
\allowdisplaybreaks

\usepackage{color}

\makeatletter
\renewcommand\section{\@startsection {section}{1}{\z@}%
  {-3ex \@plus -1ex \@minus -.2ex}%
  {2ex \@plus.1ex}%
  {\normalfont\normalsize\bf\centerline}}
\renewcommand\subsection{\@startsection{subsection}{2}{\z@}%
  {-2.5ex\@plus -0.4ex \@minus -.2ex}%
  {1.4ex \@plus .1ex}%
  {\noindent\normalfont\small\bf\centerline}}
\renewcommand\subsubsection{\@startsection{subsubsection}{3}{\z@}%
  {-0.6ex\@plus -0.2ex \@minus -.2ex}%
  {0.4ex \@plus .1ex}%
  {\normalfont\normalsize\it}}
\makeatother

\def\Xint#1{\mathchoice
   {\XXint\displaystyle\textstyle{#1}}%
   {\XXint\textstyle\scriptstyle{#1}}%
   {\XXint\scriptstyle\scriptscriptstyle{#1}}%
   {\XXint\scriptscriptstyle\scriptscriptstyle{#1}}%
   \!\int}
\def\XXint#1#2#3{{\setbox0=\hbox{$#1{#2#3}{\int}$}
     \vcenter{\hbox{$#2#3$}}\kern-.5\wd0}}

\def\dashint{\Xint-}

\begin{document}

\title{Whitham modulation theory for the two-dimensional Benjamin-Ono equation}

\author{Mark Ablowitz$^1$}
\author{Gino Biondini$^{2,3}$}
\author{Qiao Wang$^2$}

\affiliation{%
1: Department of Applied Mathematics, State University of Colorado, Boulder, CO 80303
\\
2: Department of Mathematics, State University of New York, Buffalo, NY 14260
\\
3: Department of Physics, State University of New York, Buffalo, NY 14260
}

\date{\today}

\begin{abstract}
Whitham modulation theory for the two dimensional Benjamin-Ono (2DBO) equation is presented.
A system of five quasi-linear first-order partial differential equations is derived.
The system describes modulations of the traveling wave solutions of the 2DBO equation.
These equations are 
transformed to a singularity-free hyrdodynamic-like system referred to here as the 2DBO-Whitham system.
Exact reductions of this system are discussed,
the formulation of initial value problems is considered, 
and the system is used to study the transverse stability of traveling wave solutions of the 2DBO equation.
\end{abstract}


\maketitle


\section{I.~ Introduction}

Small-dispersion limits and dispersive shock waves (DSWs) have been intensely studied during the last fifty years.  
There are numerous physical applications of DSWs in fluid dynamics, nonlinear optics, Bose-Einstein condensates, 
magnetic films and thermal media,
amongst others,
cf.~\cite{physicad,elhoefer2016physd,elgrimshaw2002,marchantsmyth,D9,Hoefer06,Wan07,Gofra07,prl2016deng,janantha}
and references therein. 
Most of the studies in the literature have been devoted to (1+1)-dimensional systems, however, 
and much less is known about multi-dimensional systems.

Following some earlier works~\cite{bogaevskii,infeldrowlands,krichever}, 
recently there has been considerable attention devoted to the study of 
small dispersion problems for
(2+1)-dimensional systems
\cite{klein,cbo,kpwhitham}. 
One of the goals of this work is to develop tools that can be used to describe the behavior of DSWs in multi-dimensional settings.

Steps forward in this direction were recently presented in~\cite{kpwhitham,cbo}.
In particular, 
the formation of DSWs along curved fronts was studied in~\cite{cbo},
and in \cite{kpwhitham} a 2D generalization of Whitham modulation theory was formulated in terms of Riemann-type variables and used to study important properties associated with
the small dispersion limit of the Kadomtsev-Petviashvili (KP) equation~\cite{KP}.

In this work we use similar methods to study the small dispersion limit 
of the two-dimensional Benjamin-Ono (2DBO) equation~\cite{2dbo},
\vspace*{-0.6ex}
\[
\label{e:2dbo}
(u_t + u u_x + \epsilon \mathcal{H} [u_{xx}])_x + \lambda u_{yy} = 0\,,
\]
where 
subscripts $x,y,t$ denote partial differentiation,
$0<\epsilon \ll1$ is a small parameter quantifying the relative strength of dispersive effects
and $\mathcal{H}$ is the Hilbert transform operator, defined by
\vspace*{-1ex}
\[
\label{e:hilberttransform}
\mathcal{H}[f(x)] = \frac{1}{\pi} ~ \dashint_{-\infty}^\infty \frac{f(y)}{y-x} dy\,,
\nonumber
\]
where $\dashint$ denotes the Cauchy principal value integral (cf.~\cite{nist}).
Equation~\eqref{e:2dbo} is a two-dimensional (2D) extension of the classical (i.e., one-dimensional) Benjamin-Ono (1DBO) equation~\cite{benjamin,ono}
\vspace*{-0.6ex}
\[
\label{e:bo}
u_t + u u_x + \epsilon \mathcal{H} [u_{xx}] = 0\,,
\]
and describes weakly nonlinear long internal waves in fluids of great depth~\cite{2dbo}.
By analogy with the KP equation, 
the cases $\lambda = -1$ and $\lambda = 1$ are referred to as the 2DBOI equation and 2DBOII equation, respectively.
Similarly to what happens for the two variants of the KP equation
(e.g., see \cite{infeldrowlands,AS1981}),
the 2DBOII equation (like the KPI equation and the second case of the two-dimensional intermediate-long wave equation, or 2DLWII) 
arises when surface tension is negligible \cite{2dbo},
whereas 
the 2DBOI equation (and the 2DILWI) arises when surface tension effects are dominant~\cite{ablowitzhaut08,ablowitzhaut}.

The small dispersion limit of the 1DBO equation [i.e., Eq.~\eqref{e:bo}]
has been studied extensively~\cite{jorge,matsuno1,matsuno2,matsuno3,dobrokhotov,miller}.
However, no general results are available for the 2DBO equation [i.e., Eq.~\eqref{e:2dbo}] to the best of our knowledge. 
A one-dimensional (1D) reduction of the 2DBO equation and its associated DSW behavior was studied recently in~\cite{cbo}, 
in which a similarity variable $\eta = x + P(y,t)$ was used to reduce Eq.~\eqref{e:2dbo} to the cylindrical Benjamin-Ono (cBO) equation, 
\vspace*{-0.4ex}
\[
\label{e:cbo}
u_t + u u_\eta + \frac{\lambda c}{1+2 \lambda c t} u+ \epsilon \mathcal{H} [u_{\eta\eta}] =0\,,
\]
as well as to the write the resulting equations in terms of Riemann variables and study
the DSW behavior 
with step-like initial data along parabolic fronts.
The study of more general initial conditions that do not admit a 1D reduction
is still an open problem. 

In this work we derive the 2D Whitham system for the 2DBO equation using the method of multiple scales (e.g., as in \cite{Luke}), 
and we simplify the resulting system of partial differential equations (PDEs) by suitably rewriting it in terms of Riemann-type variables.
We then discuss various properties of the resulting system of equations, including exact reductions
and the formulation of a 2D generalization of the Riemann problem for the 1D Whitham system.
Finally, 
we use the system to investigate the stability of the traveling wave solutions of 2DBO equation. 
Note that, unlike the KP equation, Eq.~\eqref{e:2dbo} is not known to be an integrable system. 
The methods presented here do not rely on integrability.



\section{II.~ Derivation of the 2DBO-Whitham system}
\label{s:derivation}

The derivation of the modulation equations for the 2DBO equation is similar to that for the KP equation, 
and we refer the reader to \cite{kpwhitham} for further details.


\subsection{II.A~ The multiple scales expansion}
\label{s:multiscale}

We begin by rewriting the 2DBO equation as the system
\vspace*{-0.4ex}
\bse
\label{e:evol}
\begin{align}
\label{e:2dbo1}
&u_t + u u_x + \epsilon \mathcal{H} [u_{xx}] + \lambda v_y = 0\,, \\
\label{e:2dbo2}
&v_x = u_y\,.
\end{align}
\ese
We look for solutions of Eq.~\eqref{e:2dbo1} as
$u(\theta, x, y, t)$, 
with the rapidly varying variable $\theta(x,y,t)$ defined by
\vspace*{-0.4ex}
\[
\label{e:thetaderiv}
\theta_x = {k(x,y,t)}/{\epsilon},~~
\theta_y = {l(x,y,t)}/{\epsilon},~~
\theta_t = -{\omega(x,y,t)}/{\epsilon},
\]
where $k,~l$ and $\omega$ are the local wave numbers and frequency, respectively, which are assumed to be slowly varying functions of $x$, $y$ and $t$.
Enforcing the equality of the mixed second derivatives of $\theta$ then yields three compatibility conditions
\vspace*{-0.4ex}
\bse
\label{e:compatibility}
\begin{gather}
k_t + \omega_x = 0\,,\qquad
l_t + \omega_y = 0\,,
\label{e:eq12}
\\
k_y - l_x = 0\,.
\label{e:constraint}
\end{gather}
\ese
Equations~\eqref{e:eq12} are referred to as the equations of conservation of waves,
and provide the first and second modulation equations. 
Note that Eqs.~\eqref{e:eq12} automatically imply that Eq.~\eqref{e:constraint} is satisfied for all $t>0$ if 
it is satisfied at $t =0$.

In terms of the fast variable $\theta$ and the slow variables $x,y,t$, 
Eqs.~\eqref{e:evol} become
\vspace*{-0.6ex}
\bse
\label{e:2dboasymp}
\begin{align}
&\big(-\omega u_\theta + k u u_\theta + k^2 \mathcal{H}[u_{\theta\theta}] + \lambda l v_\theta \big) /\epsilon
\nonumber\\
&\kern3em{ }
+ u_t + u u_x +\mathcal{H} \big[k_x u_\theta + 2 k u_{\theta x}\big] + \lambda v_y 
\nonumber\\
&\kern14em{ }
  +\epsilon \mathcal{H} \big[u_{xx}\big] = 0\,, 
\\
&\big( k v_\theta - l u_\theta \big)/\epsilon + \big( v_x - u_y \big) = 0\,.
\end{align}
\ese
We then look for a perturbative solution for $\@u = (u,v)^T$ as
\vspace*{-0.6ex}
\begin{gather}
\@u = \@u^{(0)}(\theta, x,y,t) + \epsilon \@u^{(1)}(\theta, x,y,t) + O(\epsilon^2)\,.
\label{e:uvasymp}
\end{gather}
Substituting Eqs.~\eqref{e:uvasymp} into Eqs.~\eqref{e:2dboasymp} and collecting terms in the same power of $\epsilon$, one obtains 
a sequence of equations.
%
The leading-order terms, at $O(1/\epsilon)$, yield
\bse
\label{e:leading}
\begin{gather}
\label{e:leading1}
-\omega u^{(0)}_\theta + k u^{(0)} u^{(0)}_\theta + k^2 \mathcal{H}[u^{(0)}_{\theta\theta}] + \lambda l v^{(0)}_\theta = 0\,, \\
\label{e:leading2}
k v^{(0)}_\theta - l u^{(0)}_\theta = 0\,.
\end{gather}
\ese
Equations~\eqref{e:leading} can be written in vector form as
$\@M_0~\@u^{(0)} = \@0$,
where 
$\@M_0 = \@M\,\partial_\theta$,
with
\vspace*{-0.6ex}
\begin{gather}
\@M = \begin{pmatrix} 
  \mathcal{L}     & \lambda qk\\ 
  \lambda qk & -\lambda k
\end{pmatrix}\,,
\end{gather}
$\mathcal{L} = -\omega + k u^{(0)} + k^2 \mathcal{H}[ \partial_\theta]$,
and where we defined
\vspace*{-0.6ex}
\begin{align}
q(x,y,t) = l/k\,, 
\label{e:qdef}
\end{align}
which will play an important role in the following.
Integrating Eq.~\eqref{e:leading2} with respect to $\theta$, we obtain 
\vspace*{-0.6ex}
\[
\label{e:v0def}
v^{(0)} = q u^{(0)} + p\,,
\] 
where $p(x,y,t)$ is to be determined at higher order in the expansion.
Next we look at $O(1)$ terms, which yield
$\@M_1~\@u^{(1)} = \@G[\@u^{(0)}]$,
where
$\@G[\@u] = (g_1,g_2)^T$ 
and 
$\@M_1 = \partial_\theta \@M$, 
and with
\bse
\label{e:Gdef}
\begin{align}
  g_1[\@u] &= - u_t - u u_x - \mathcal{H} [k_x u_\theta + 2 k u_{\theta x} ] - \lambda v_y \,,\\
  g_2[\@u] & = \lambda ( v_x - u_y \big) \,. 
\end{align}
\ese
Note that $\@M_1$ is a total derivative in $\theta$, 
and the solution of Eqs.~\eqref{e:leading} is periodic, with period $P$ computed explicitly below. 
To avoid secular terms, one needs the following two-component periodicity condition:
\vspace*{-0.6ex}
\[
\label{e:eq34}
\int_{0}^{P} \@G[\@u^{(0)}] ~d\theta = 0\,,
\]
which provides two further modulation equations.
Finally, the Fredholm solvability condition for the inhomogeneous problem at $O(1)$ yields the last modulation equation:
\vspace*{-0.6ex}
\[
\label{e:eq5}
\int_{0}^{P} \@u^{(0)} \cdot \@G[\@u^{(0)}] ~d\theta = 0\,.
\]
Equations~\eqref{e:eq12}, \eqref{e:eq34} and~\eqref{e:eq5} comprise the system of five modulation equations.


\subsection{II.B~ Leading-order solution and modulation equations}
\label{s:elliptic}

We now write PDEs for the evolution of the characteristic parameters of the traveling wave solutions of the 2DBO equation.
We return to the equations at leading order and use
Eq.~\eqref{e:leading2} to rewrite Eq.~\eqref{e:leading1} as
\vspace*{-0.6ex}
\[
\label{e:genus1eqn}
k \mathcal{H}[u^{(0)}_{\theta\theta}] + u^{(0)} u_\theta^{(0)} -V u_\theta^{(0)}  = 0\,,
\]
where 
\begin{equation}
\label{e:Omega}
V + \lambda q^2 = {\omega}/{k} = \Omega \,.
\end{equation}
The solution of Eq.~\eqref{e:genus1eqn} is~\cite{benjamin} 
\[
\label{e:genus1soln}
u^{(0)}(\theta,x,y,t) = \frac{4 k^2}{\sqrt{A^2 + 4 k^2} - A \cos (\theta - \theta_0)} + \beta\,,
\]
where $\theta_0$ is a constant and the phase velocity $V$ is given by
\vspace*{-0.6ex}
\[
\label{e:Vdef2}
V = (1/2)\sqrt{A^2 + 4 k^2} + \beta\,.
\]
Unlike the periodic solutions of the KP equation, 
the solution~\eqref{e:genus1eqn} involves trigonometric (as opposed to elliptic) functions; 
its period as a function of $\theta$ is simply $P = 2 \pi$.
When $k,~V,~\beta$ and $q$ are constants, Eq.~\eqref{e:genus1soln} is a 2D extension of the periodic solution of the 1DBO equation~\cite{benjamin}. 
When these quantities are slowly varying functions of $x$, $y$ and $t$, Eq.~\eqref{e:genus1soln} describes a slowly modulated periodic wave,
whose evolution is determined by the five modulation equations above.

Substituting Eqs.~\eqref{e:Gdef} into Eqs.~\eqref{e:eq34} and~\eqref{e:eq5} we have
\bse
\begin{align}
\label{e:meq3}
&\partialderiv{G_1}{t} + \frac{1}{2} \partialderiv{G_2}{x} + \lambda \partialderiv{ }y \big( q\,G_1+ 2 \pi p\big) = 0\,,
\\
\label{e:meq4}
&\partialderiv{G_2}{t} + \frac{2}{3} \partialderiv{G_3}{x} + 2 G_4
\nonumber\\
&\kern1em{ }
  + \lambda\bigg[2 G_2 \DDy q + 2 q \partialderiv{G_2}{y} - q^2 \partialderiv{G_2}{x}+2 G_1 \DDy p \bigg] = 0\,, 
\\[-1ex]
\label{e:meq5} 
&
\partialderiv px + \frac1{2\pi} G_1 \partialderiv qx - \frac1{2\pi}\DDy{G_1} = 0\,,
\end{align}
where
\begin{align*}
&G_j = \int_{0}^{2\pi} (u^{(0)})^j ~d\theta\,,\qquad j=1,2,3,\\
&G_4 = \int_{0}^{2\pi} u^{(0)} \mathcal{H} [k_x u^{(0)}_\theta + 2 k u^{(0)}_{\theta x} ] ~d\theta\,,
\end{align*}
\ese
and where we introduced the ``convective'' derivative
\[
\label{e:DDy}
\frac{D}{D y} = \partialderiv{ }y - q \partialderiv{ }x\,.
\]
Specifically, Eq.~\eqref{e:meq4} follows from the Fredholm solvability condition~\eqref{e:eq5},
Eq.~\eqref{e:meq3} from the first component of the periodicity condition~\eqref{e:eq34}
and Eq.~\eqref{e:meq5} from the second component of Eq.~\eqref{e:eq34}, which is simply
\[ 
\int_0^{2\pi}(v_x^{(0)}-u_y^{(0)})\, d \theta=0\,.
\]
Explicitly, using Eq.~\eqref{e:genus1soln}, we have
\vspace*{-1ex}
\bse
\label{e:integrals}
\begin{align}
G_1 &= 2 \pi (\beta + 2 k)\,,\\
G_2 &= 2 \pi (\beta^2 + 4 V k)\,, \\
G_3 & = 2 \pi (\beta^3 - 6 \beta^2 k + 12 k \beta V + 3 k A^2 + 8 k^3 )\,, \\
G_4 &= - \pi (k A^2)_x\,.
\end{align}
\ese
Thus, Eqs.~\eqref{e:eq12}, \eqref{e:meq3}, \eqref{e:meq4} and~\eqref{e:meq5} become
\bse
\label{e:meqns}
\begin{align}
\label{e:dkdt}
&k_t +(\Omega k )_x=0\,,
\\
\label{e:dqdt_nondiag}
&(k q)_t +(\Omega k )_y =0\,, 
\\
\label{e:nondiagmodeq1}
&(\beta + 2 k)_t + \frac12(\beta^2 + 4 V k)_x  + \lambda \big[ q (\beta + 2 k) + p]_y = 0 \,,
\\
\label{e:nondiagmodeq2}
&(\beta^2 + 4 V k)_t + \frac23 (\beta^3 + 6 k V^2 + 2 k^3)_x + \lambda \bigg\{ 2 \DDy p (\beta + 2 k)
\nonumber\\
&\kern1em{ } 
 + \Big[2 \DDy q + 2 q \partial_y - q^2 \partial_x \Big] (\beta^2 + 4 V k)  \bigg\} = 0\,,
\\
\label{e:pconstraint}
&\big[ q (\beta + 2 k) + p]_x - (\beta + 2 k)_y  = 0\,.
\end{align}
\ese
Equations~\eqref{e:meqns} comprise four evolution PDEs and one non-evolutionary constraint [Eq.~\eqref{e:pconstraint}]
for the five dependent variables $V$, $\beta$, $k$, $p$ and $q$.
When $p$ is a constant, $q$ vanishes identically
and the other dependent variables are independent of $y$, 
Eqs.~\eqref{e:meqns} reduce to the modulation equations for the 1DBO equation~\cite{matsuno1}.


\subsection{II.C~ Transformation to Riemann-type variables}
\label{s:riemannvariables}

Next we introduce Riemann-type variables to simplify the system~\eqref{e:meqns}.
Namely, we define the variables $r_1$, $r_2$ and $r_3$
as in the Riemann invariants for the 1DBO equation
by letting \cite{matsuno1,matsuno2,cbo},
\[
\label{e:riemannvariables}
V = r_2 + r_3\,, \quad k = r_3 - r_2\,, \quad \beta = 2 r_1\,.
\]
(This transformation is similar to the one for the Korteweg-de Vries equation~\cite{Whitham}, 
and $r_1,r_2,r_3$ 
are obtained from $V$ and $k$ and $\beta$ by inverting Eqs.~\eqref{e:riemannvariables}.)
\,\ 
In terms of $r_1,\,r_2,\,r_3$ and $q$, the leading-order solution of the 2DBO equation is 
\vspace*{-0.6ex}
\begin{multline}
\label{e:periodicsoln}
u^{(0)}(x,y,t) = 2 r_1 + 2 (r_3-r_2)^2/
\\
\bigg[(r_3+r_2-2 r_1) - 2 \sqrt{(r_2-r_1)(r_3-r_1)} \cos\theta\bigg]\,,
\end{multline}
with $\theta$ determined (up to an integration constant) by Eqs.~\eqref{e:thetaderiv}.
When $r_2\to r_1$, Eq.~\eqref{e:periodicsoln} reduces to a constant.
When $r_2\to r_3$, Eq.~\eqref{e:periodicsoln} yields the line-soliton solutions of Eq.~\eqref{e:2dbo}:
\vspace*{-1ex}
\[
u(x,y,t) = 2 r_1 + \frac{8 (r_3-r_1)}{4(r_3-r_1)^2[x+q y-(2 r_3+\lambda q^2)t]^2 + 1}\,.
\label{e:bosoliton}
\]
(Note however that the solution in Eq.~\eqref{e:bosoliton} decays algebraically as $x\to\pm\infty$, unlike the line solitons of the KP equation.)

Rewriting the system~\eqref{e:meqns} in terms of 
$\@ = (r_1,~r_2,~r_3,~q,~p)^T$, one obtains the hydrodynamic system
$R ~\textbf{r}_t + S ~\textbf{r}_x +T ~\textbf{r}_y  = 0$,
where 
$R$, $S$ and $T$ are $5\times5$ real-valued matrices.
In particular, $R$ has block structure
$R = \diag(R_4,0)$, 
where $R_4$ denotes a $4\times4$ matrix.
Even though $R$ is not invertible, we can multiply the vector equations from the left by the ``pseudo-inverse''
$\~R^{-1} = \diag(R_4^{-1},0)$,
obtaining
\vspace*{-0.6ex}
\[
\label{e:matrixeqn}
I \textbf{r}_t + A~\textbf{r}_x +B~\textbf{r}_y = 0\,,
\] 
where $I = \mathrm{diag}(1,1,1,1,0)$, $A = R^{-1}S$ and $B = R^{-1}T$.

While the Whitham system for the 1DBO equation is diagonalized by the above transformation to Riemann variables,
one cannot find a change of dependent variables to diagonalize the corresponding system~\eqref{e:matrixeqn} for the 2DBO equation,
since $AB \neq BA$.




The system~\eqref{e:matrixeqn} becomes singular as $r_2\to r_1$ and as $r_2\to r_3$.
That is, some entries of both $A$ and $B$ become infinite in these limits.
(These singularities do not arise in the Whitham systems for the 1DBO and cBO equations,
and occur even though the determinants and eigenvalues of $A$ and $B$ remain finite.)
\,\ 
However, one can obtain an equivalent but simplified system 
that is free of singularities, as we show next.

Using the definition~\eqref{e:riemannvariables} and $q = l/k$, 
the third compatibility condition, Eq.~\eqref{e:constraint}, can be written as
\vspace*{-0.6ex}
\[
\frac{Dr_3}{Dy} - \frac{Dr_2}{Dy} - (r_3 - r_2) \frac{\partial q}{\partial x} = 0\,.
\label{e:constr}
\]
Equation~\eqref{e:constr} is identically satisfied when $q$ is zero and $r_1,r_2,r_3$ are independent of $y$.
Subtracting a suitable multiple of the constraint~\eqref{e:constr} from each equation, and a suitable multiple of Eq.~\eqref{e:pconstraint} from the other equations,
the five modulation equations take on the particularly simple form,
which is also completely free of singularities: 
\bse
\label{e:2dbowhitham}
\begin{align}
&\frac{\partial r_j}{\partial t} + (V_j + \lambda q^2 ) \frac{\partial r_j}{\partial x} 
+ 2 \lambda q \frac{D r_j}{D y} + \lambda \nu_j \frac{D q}{D y} + \frac{\lambda}{2} \DDy{p} = 0\,, \nonumber \\ 
&\kern16em{ } \quad j = 1,2,3, \\
\label{e:q}
&\frac{\partial q}{\partial t} + (V_2 + \lambda q^2 ) \frac{\partial q}{\partial x}
+ 2 \lambda q \frac{D q}{D y} + 2 \frac{D r_3}{D y}= 0\,, \\
\label{e:pconstr}
&\partialderiv{p}{x} - 2 \DDy{r_1} + 2 r_1 \partialderiv{q}{x} = 0\,,
\end{align}
\ese
where 
\vspace*{-1ex}
\bse
\label{e:vnudef}
\begin{gather}
V_j = 2 r_j\,,\qquad j = 1,2,3,
\\
\nu_1 = r_3 - r_2 + r_1\,,\qquad 
\nu_2 = \nu_3 = r_3 + r_2 - r_1\,.
\end{gather}
\ese
That is, all coefficients in Eqs.~\eqref{e:2dbowhitham} have finite limit as $r_2 \to r_1$ and as $r_2 \to r_3$.
Note that $V_1,\dots,V_3$ are exactly the characteristic speeds of the 1DBO-Whitham system.
Also, $\nu_1,\dots,\nu_3$ are exactly the same as the coefficients appearing in the inhomogeneous terms for the cBO-Whitham system~\cite{cbo}.
\textit{Hereafter we refer to Eqs.~\eqref{e:2dbowhitham} as the 2DBO-Whitham system.}


\subsection{II.D~ General nature of the last two modulation equations}

One of the novelties of the system~\eqref{e:2dbowhitham} compared to the 1D case is 
the presence of Eqs.~\eqref{e:q} and~\eqref{e:pconstr}, which determine the new dependent variables $q$ and~$p$.
An alternative but equivalent version of these two equations 
can be obtained by noting that they 
can be derived separately from the equations for the Riemann-type variables $r_1,r_2,r_3$
in a straightforward way. 

Indeed, using the first of Eqs.~\eqref{e:compatibility}, \eqref{e:qdef} and~\eqref{e:Omega},
the second of Eq.~\eqref{e:compatibility} yields
\begin{equation}
\label{e:dqdtnew}
\partialderiv qt + \Omega \partialderiv qx +\DDy{\Omega} = 0\,,
\end{equation}
where $\Omega = \omega/k = V + \lambda q^2$ as before.
Importantly, 
\textit{Eq.~\eqref{e:dqdtnew} arises whenever one seeks multiple-scale solutions of a multi-dimensional system, 
leading to the compatibility conditions~\eqref{e:compatibility}.}
Thus,
the only difference between the 2DBO Eq.~\eqref{e:2dbo} and other evolution equations is just 
how $\Omega$ is given in terms of the other dependent variables.
For example, for the KP equation one also has $\Omega = V + \lambda q^2$,
but $V = 2(r_1 + r_2 + r_3)$ in that case, 
whereas for the 2DBO equation we have $V = r_2 + r_3$.

Similarly, the constraint for $p$, namely Eq.~\eqref{e:meq5},
also takes essentially the same form as 
for the KP equation.
The only difference is
how $G_1$ depends on the Riemann-type variables.
For the 2DBO equation we have 
$G_1=4\pi(r_1+r_3-r_2)$,
whereas the expression for the KP equation is slightly more complicated.

Substituting $\Omega$ and $G_1$ in equations~\eqref{e:dqdtnew} and~\eqref{e:meq5} yields, respectively
\bse
\label{e:pqnew}
\begin{gather}
\label{newqeqA}
\partialderiv qt + (r_2+r_3+\lambda q^2)\partialderiv qx + \DDy{ }(r_2+r_3+\lambda q^2) = 0\,,
\\
\label{newpeqA}
\partialderiv px + 2(r_1+r_3-r_2)\partialderiv qx - 2\DDy{ }(r_1+r_3-r_2) = 0\,.
\end{gather}
\ese
These equations can be transformed to Eqs.~\eqref{e:q} and~\eqref{e:pconstr}
using the compatibility condition~\eqref{e:constraint}.  Also,
the equations for $r_1,r_2,r_3$ 
can be obtained by ``diagonalizing'' 
Eqs.~\eqref{e:dkdt}, \eqref{e:nondiagmodeq1} and~\eqref{e:nondiagmodeq2},
which are the analogue of the modulation equations for the 1DBO.
For brevity we omit the details, and we will report on these issues in a future publication.


\section{III.~ Reductions and Riemann problems}
\label{s:properties}

We now discuss reductions of the 2DBO-Whitham system
as well as the choice of initial conditions (ICs) and boundary conditions (BCs) to obtain well-posed initial value problems (IVPs) for it,
including generalizations of the Riemann problem for the 1DBO-Whitham system~\cite{matsuno2}.


\subsection{III.A~ Exact reductions of the 2DBO-Whitham system}
\label{s:reductions}


\paragraph{One-dimensional reductions.}
Suppose that $r_1,r_2,r_3$ depend on $x$ and $y$ only through the similarity variable $\eta = x+P(y,t)$
and $q = P_y(y,t)$, with $p(x,y,t)$ a constant. 
Then it is straightforward to see that $Dr_j/Dy =0$.
Since $q$ is independent of $x$ and $p$ is a constant, Eqs.~\eqref{e:2dbowhitham} become 
\vspace*{-1ex}
\bse
\begin{align}
\label{e:first3eqn}
&\frac{\partial r_j}{\partial t} + P_t \frac{\partial r_j}{\partial \eta}+ (V_j + \lambda P_y^2)\,\frac{\partial r_j}{\partial \eta}
+ \lambda \nu_j P_{yy} = 0\,,~~~ j=1,2,3,
\\[-1ex]
\label{e:qeqn}
&\frac{\partial q}{\partial t} + 2 \lambda q \frac{\partial q}{\partial y} = 0\,.
\end{align}
\ese
(The fifth modulation equation is identically satisfied in this case.)1
In terms of $P_y$, Eq.~\eqref{e:qeqn} is $P_{ty} + 2 \lambda P_y P_{yy} = 0$, which after integration yields
$P_t + \lambda P_y^2 = 0$.
In turn, the system of equations~\eqref{e:first3eqn} becomes
\vspace*{-1ex}
\begin{align}
\label{e:cboreduction}
\frac{\partial r_j}{\partial t} + V_j\,\frac{\partial r_j}{\partial \eta} + \lambda \nu_j P_{yy} = 0\,, \qquad j = 1,2,3\,.
\end{align}
In order for this setting to be self-consistent, however, the last term in the LHS of Eqs.~\eqref{e:cboreduction} must be independent of~$y$.
Therefore, only three possibilities arise:

(i) $P_y = 0$, in which case one simply has $q(x,y,t)=0$ (implying that the resulting behavior is one-dimensional)
and $P(y,t) = 0$, as well as $\eta = x$. In this case, the system~\eqref{e:cboreduction} reduces to the Whitham system for the 1DBO equation~\cite{matsuno1}.

(ii) $P_y = a$ is a nonzero constant, in which case one has $q(x,y,t)=a$ and $P(y,t) = ay$, implying $\eta = x + ay$. 
Then system~\eqref{e:cboreduction} reduces to the 1D Whitham system 
with $x$ replaced by $\eta$.

(iii) $P_{yy} = f(t)$ is a function of $t$,  in which case $q = P_y = f(t) y$, 
and Eq.~\eqref{e:qeqn} now yields 
$f_t + 2 \lambda f^2=0$.
This ordinary differential equation is easily solved.
In particular,
for a constant IC $f(0)=c={}$const, we have
$f(t) = {c}/({1+2c\lambda t})$,
and hence
$q(y,t) = {cy}/({1+2c\lambda t})$, 
which reduces system~\eqref{e:cboreduction} to the Whitham system for the cBO equation~\cite{cbo}.

\paragraph{Genus-zero reductions.}
Two further exact reductions of the system~\eqref{e:2dbowhitham} 
are obtained when $r_1 = r_2$ and $r_2=r_3$, respectively. 
In the first case, the leading-order periodic solution degenerates to a constant with respect to the fast variable, 
while the second one yields the solitonic limit.

When $r_1=r_2$, the PDEs for $r_1$ and $r_2$ coincide.
As a result, system~\eqref{e:2dbowhitham} reduces to the following $4\times4$ system:
\bse
\label{e:2dbowhitham1reduction}
\begin{align}
&\partialderiv{r_1}{t} + (2 r_1 + \lambda q^2)\,\partialderiv{r_1}{x} + 2 \lambda q  \DDy{r_1} + \lambda r_3 \DDy{q} + \frac{\lambda}{2} \DDy{p} = 0\,,
\\
&\partialderiv{r_3}{t} + (2 r_3 + \lambda q^2)\,\partialderiv{r_3}{x} + 2 \lambda q  \DDy{r_3} + \lambda r_3 \DDy{q} + \frac{\lambda}{2} \DDy{p} = 0\,,
\\
&\partialderiv{q}{t} + (2 r_1 + \lambda q^2)\,\partialderiv{q}{x} + 2 \lambda q \DDy{q} 
 + 2 \DDy{r_3} = 0\,, 
\\
&\partialderiv {p}{x} - 2 \DDy{r_1} + 2 r_1 \partialderiv {q}x = 0\,.
\end{align}
\ese
Similarly, when $r_2=r_3$, the PDEs for $r_2$ and $r_3$ coincide, and 
system~\eqref{e:2dbowhitham} reduces to
\bse
\label{e:2dbowhitham2reduction}
\begin{align}
&\partialderiv{r_1}{t} + (2 r_1 + \lambda q^2)\,\partialderiv{r_1}{x} + 2 \lambda q  \DDy{r_1} + \lambda r_1 \DDy{q} + \frac{\lambda}{2} \DDy{p} = 0\,,
\\
&\partialderiv{r_3}{t} + (2 r_3 + \lambda q^2)\,\partialderiv{r_3}{x} + 2 \lambda q  \DDy{r_3}
\nonumber \\[-0.4ex]
&\kern6em{ }
+ \lambda (2 r_3 - r_1) \DDy{q} 
+ \frac{\lambda}{2} \DDy{p} = 0\,,
\\
&\partialderiv{q}{t} + (2 r_3 + \lambda q^2)\,\partialderiv{q}{x} + 2 \lambda q \DDy{q} 
 + 2 \DDy{r_3} = 0\,, 
\\
&\partialderiv {p}{x} - 2 \DDy{r_1} + 2 r_1 \partialderiv {q}x = 0\,.
\end{align}
\ese


\subsection{III.B~ Initial-value problems for the 2DBO-Whitham system}
\label{s:IVP}


\paragraph{ICs.}

The problem of mapping an IC for $u$ to ICs for the variables $r_1,r_2,r_3$ is the same as in the 1DBO case.
Once this step is done, one can determine the IC for $q$ using the constraint~\eqref{e:constraint} at $t=0$, obtaining 
$k_y(x,y,0) = [k(x,y,0) q(x,y,0)]_x$.
Note that $k(x,y,0)$ is easily found using Eqs.~\eqref{e:riemannvariables}.
Integrating this equation, we then have
\vspace*{-0.6ex}
\begin{align}
&q(x,y,0) = \frac1{k(x,y,0)}\bigg( C_0 + \int_{x_0}^x k_y(\xi,y,0) d \xi \bigg)
\label{e:q-ic}
\end{align}
where 
$C_0= q(x_0,y,0)k(x_0,y,0)$
and $k(x,y,0)$ is assumed to be non-zero.
Also, integrating Eq.~\eqref{e:pconstr} determines $p$ for all $t\ge0$ up to an arbitrary function of $y$ and $t$, that is
%
\vspace*{-1ex}
\begin{gather}
p(x,y,t) = p^-(y,t) + \partial_x^{-1} \bigg[ 
  2 \DDy{r_1}  - 2 r_1 \partialderiv{q}{x} 
  \bigg]\,, 
\label{e:pintegrated}
\\
\noalign{\noindent where}
\label{e:dxinverse}
\partial_x^{-1}[f] = \int_{-\infty}^x f(\xi,y,t)\,\d\xi\,. 
\end{gather}
%
Hence, the problem is reduced to the choice of suitable BCs.

\paragraph{BCs.}
%
In the Riemann problem for the 1DBO~\cite{matsuno2}, the asymptotic values of $r_1,r_2,r_3$ as $x\to\pm\infty$ are constants.
Already in the Riemann problem for the cBO equation, however, this is not true anymore,
and the BCs for $r_j$ can be obtained from solving a reduced system of ODEs for $t$~\cite{cbo}. 
In the full 2DBO-Whitham system,
the BCs of the Riemann invariants may in general also depend on the independent variable $y$.

To make the above discussion more precise, we first go back to the 2DBO equation.
Integrating Eq.~\eqref{e:2dbo2} yields
\begin{equation}
v(x,y,t) = v^-(y,t) + \partial_x^{-1} [ u_y ]\,,
\label{e:vintegrated}
\end{equation}
where we use the superscript ``$-$'' to indicate limiting values as $x\to-\infty$, 
and $\partial_x^{-1}$ is defined by Eq.~\eqref{e:dxinverse} as before.
Substituting Eq.~\eqref{e:vintegrated} into Eq.~\eqref{e:2dbo1} yields 
\begin{equation}
u_t + uu_x + \epsilon \mathcal{H} [u_{xx}] + \lambda\,\partial_x^{-1}[u_{yy}] + \lambda\,\partial_y v^- = 0\,.
\label{e:2dboevol}
\end{equation}
Taking the limit of Eq.~\eqref{e:2dboevol} as $x\to-\infty$ we see that, if one is interested in solutions~$u$ 
which tend to constant values as $x\to-\infty$ (i.e., $u^-$ independent of $t$),
one needs $\partial_y v^-(y,t) = 0$.
Ignoring an unnecessary function of time, we can then take $v^-(y,t) = 0$.
And Eq.~\eqref{e:v0def} then leads to 
\begin{equation}
p^- + u^- q^- = 0\,,
\label{e:pBC}
\end{equation}
which determines $p^-$.
Similar arguments carry over to the 2DBO-Whitham system.
%
That is, taking the limit $x\to-\infty$, the system~\eqref{e:2dbowhitham} becomes
\vspace*{-1ex}
\bse
\label{e:BCs}
\begin{align}
\label{e:BC1}
&\partialderiv{r_j^-\!}{t} + 2 \lambda q^-  \ddy{r_j^-} + \lambda \nu_j^- \ddy{q^-} + \frac{\lambda}{2} \ddy{p^-}
  = 0\,, \quad j=1,2,3,
\\[-1ex]
\label{e:BC2}
&\partialderiv{q^-\!}{t} + 2 \lambda q^- \ddy{q^-} + 2 \ddy{r_3^-} = 0\,,  
\\
\noalign{\noindent
which determine the time evolution of $r_1^-,r_2^-,r_3^-$ and $q^-$, plus}
&\partial{r_1^-}/\partial y = 0\,,
\label{e:BC3}
\end{align}
\ese
which would seem to impose a limitation on the admissible BCs.
We next show, however, that when $r_1^- = r_2^-$ or $r_2^- = r_3^-$, 
this condition on $r_1$ is satisfied automatically, 
making Eqs.~\eqref{e:BCs} a self-consistent system.

When $r_1^-=r_2^-$
one has $u^- = 2 r_3^-$.
Hence, Eqs.~\eqref{e:BC1} with $j=1$ and Eq.~\eqref{e:BC3} yield $\partial r_1^-/\partial y = 0$ 
and $\partial r_1^-/\partial t = 0$.
Moreover, the PDEs obtained from Eqs.~\eqref{e:BC1} with $j=1,2$ coincide (as they should, since $r_1^- = r_2^-$).
Finally, Eqs.~\eqref{e:BC1} with $j=3$ is identically satisfied since $u^- = r_3^-$ is a constant,
and Eq.~\eqref{e:BC2} yields an (1+1)-dimensional Hopf equation which  
determines the time evolution of $q^-$:
\vspace*{-1ex}
\begin{equation}
\label{e:2dbowhithamleft3}
\partialderiv{q^-\!}t + 2 \lambda q^- \partialderiv{q^-\!}y = 0\,.
\end{equation}

Similarly, when $r_2^-=r_3^-$, one has $u^- = 2 r_1^-$. 
Hence, Eqs.~\eqref{e:BC1} with $j=1$ and Eq.~\eqref{e:BC3} yield $\partial r_1^-/\partial y = 0$ 
and $\partial r_1^-/\partial t = 0$.
Moreover, 
the PDEs obtained from Eqs.~\eqref{e:BC1} with $j=2,3$ coincide (as they should, since $r_2^- = r_3^-$).
Finally, Eqs.~\eqref{e:BC1} with $j=3$ and Eq.~\eqref{e:BC2} yield the following system of 2 (1+1)-dimensional ODEs for
$r^- = r_3^-$ and $q^-$:
\bse
\label{e:2dbowhithamright3}
\begin{align}
&\partialderiv{r^-\!}{t} + 2 \lambda q^- \partialderiv{r^-\!}{y} + \lambda (2 r^- - u^-)\partialderiv{q^-}{y}= 0\,,
\\
&\partialderiv{q^-}{t} + 2 \lambda q^- \partialderiv{q^-}{y} + 2 \partialderiv{r^-}{y} = 0\,,
\end{align}
\ese

Similar considerations also apply for the BCs as $x\to\infty$.
That is, Eqs.~\eqref{e:2dbowhithamleft3} or~\eqref{e:2dbowhithamright3} 
hold as $x\to\infty$ when $r^-$ and $q^-$ are replaced by $r^+$ and $q^+$.


\paragraph{Riemann problems.}
As a special case of the above IVP,
one obtains 2D generalizations of the Riemann problem for the 1DBO equation.
More precisely, one looks for solutions of the 2DBO-Whitham system~\eqref{e:2dbowhitham} with step-like ICs corresponding to a single front:
\vspace*{-1ex}
\begin{align}
\label{e:frontic}
u(x,y,0) = \begin{cases}
1, \quad x + c(y)<0\,, \\
0, \quad x + c(y) \geq 0\,,
\end{cases}
\end{align}
with $c(y)$ arbitrary. 
As in the 1D case, one can regularize the jump by choosing
the ICs for the Riemann variables to be 
\be
\label{e:r-ic}
r_1(x,y,0) = 0,~
r_2(x,y,0) = R_2(x+c(y)),~
r_3(x,y,0) = {\textstyle\frac12},
\ee
where $R_2(\xi)$ smooths out the jump between 0 and 1/2, e.g., 
$R_2(\xi) = \frac14 \big(1+\tanh\big[\xi/\delta\big]\big)$,
with $\delta$ a small parameter.
To determine the corresponding IC for $q$,
note that Eqs.~\eqref{e:r-ic} imply that the constraint~\eqref{e:constraint} is satisfied at $t=0$.
Also, Eqs.~\eqref{e:riemannvariables} and~\eqref{e:r-ic} imply
$k(x,y,0) = {\textstyle\frac12} - r_2(x,y,0)$,
and it is easy to check that $k_y(x,y,0) = c'(y) k_x(x,y,0)$.
Therefore, substituting in Eq.~\eqref{e:q-ic}, the IC for $q$ 
reduces to
$q(x,y,0) = c'(y)$.

If $c(y)$ is constant or linear in $y$ the IC for $q$ is trivial, whereas if $c(y)$ is a quadratic function of $y$ one reduces to the ICs of the Riemann problem for the cBO equation.
Finally, the IC for $p$ is chosen as described earlier, namely via Eq.~\eqref{e:pintegrated} at $t=0$ and Eq.~\eqref{e:pBC}. 


\section{IV.~ Stability of periodic solutions} 
\label{s:stability}

We now use the 2DBO-Whitham system~\eqref{e:2dbowhitham} to investigate the stability of the periodic solutions of the 2DBO equation. 

\smallskip
\paragraph{Stability analysis.}
Constant 
values of $r_1,r_2,r_3,q,p$
yield exact periodic solutions of the 2DBO equation.
To study their spectral stability, we can use 
the 2DBO-Whitham system~\eqref{e:2dbowhitham} to 
study the evolution of small initial perturbations of these constant values.
That is, we look for
\be
\label{e:perturb}
r_j = \~r_j + r_j'\,,\quad j=1,2,3,\qquad 
q = q'\,, ~~ p = p'\,,
\ee
where $\~r_1,\~r_2,\~r_3$ are arbitrary constants satisfying $\~r_1 \leq \~r_2 \leq \~r_3$,
together with 
$|r_j'(x,y,t)| \ll 1$ for $j = 1,2,3,$, $|q'(x,y,t)| \ll 1$ and $|p'(x,y,t)| \ll 1$.
Substituting Eqs.~\eqref{e:perturb} into Eqs.~\eqref{e:2dbowhitham} and dropping higher-order terms, we obtain
\vspace*{-1ex}
\begin{align*}
&\frac{\partial r_j'}{\partial t} + \~V_j \,\frac{\partial r_j'}{\partial x} + \lambda \~\nu_j \frac{\partial q'}{\partial y} 
+ \frac{\lambda}{2} \frac{\partial p'}{\partial y} = 0\,,\quad j=1,2,3,\\
&\frac{\partial q'}{\partial t} + \~V_2\,\frac{\partial q'}{\partial x}
+ 2 \frac{\partial r_3'}{\partial y} = 0\,,\quad
\frac{\partial p'}{\partial x} - 2 \frac{\partial r_1'}{\partial y} + 2 \~r_1 \frac{\partial q'}{\partial x} = 0\,,
\end{align*}
where $\~V_1,\dots,\~V_3$ and $\~\nu_1,\dots,\~\nu_3$ denote the unperturbed values
of all the corresponding coefficients, as defined in Eqs.~\eqref{e:vnudef}.
Next we look for plane wave solution of the above system of linear PDEs in the form 
\vspace*{-1ex}
\bse
\label{e:planewave}
\begin{align}
&r_j'(x,y,t) = R_j \,\e^{i(K x + L y - W t)}\,,\quad j = 1,2,3,
\\
&(q'(x,y,t),p'(x,y,t)) = (Q,P)\,\e^{i(K x + L y - W t)}\,.
\end{align}
\ese
obtaining the homogeneous linear system
\vspace*{-1ex}
\bse
\label{e:linearterm2}
\begin{align}
&(W - K \~V_j) R_j = \lambda L \~\nu_j Q + \lambda L P/2\,,\quad j=1,2,3, \\
&(W - K \~V_2) Q = 2 L R_3 \,, \quad
K P = 2 L R_1 - 2 K \~r_1 Q \,.
\end{align}
\ese
Non-trivial values of $(R_1,R_2,R_3,Q,P)$ exist if the determinant of the corresponding coefficient matrix vanishes,
which yields the linear dispersion relation
$P_4(K,L,W) = 0$,
where $P_4(K,L,W)$ is a polynomial that is cubic in $W$ and quartic in $K$ and $L$.
The periodic solution of the 2DBO equation corresponding to $\~r_1,\~r_2,\~r_3$ is therefore linearly stable 
if $W \in\Real$ for all $K,L\in\Real$, because in this case perturbations remain bounded.
Conversely, if there exist solutions with $\Im W\ne0$ for $K,L\in\Real$, 
some perturbations will grow exponentially, and the periodic solution is unstable.

In particular, for $K = 0$
(corresponding to perturbations independent of~$x$),
the dispersion relation simplifies to
\vspace*{-0.6ex}
\begin{align}
\label{e:dispersionrelation2}
(W/L)^2 = \lambda f(\~r_1,\~r_2,\~r_3)\,,
\end{align}
where $f(r_1,r_2,r_3) = 4(r_2 - r_1)$.
Since $f(r_1,r_2,r_3)$ is always non-negative for $r_1 \leq r_2$, 
for the 2DBOI equation ($\lambda = -1$) $W$ is purely imaginary, and therefore all its periodic waves are linearly unstable. 
Conversely, for the 2DBOII equation ($\lambda = 1$), $W$ is real, and therefore all its periodic waves are linearly stable 
in the spectral sense.

In the special case $r_1 = 0, r_2 = m$ and $r_3 = 1/2$, which is relevant to the Riemann problem discussed earlier,
we simply have $f(r_1,r_2,r_3) = f(m) = 4 m$.
Interestingly, the growth rate $g(m) = \sqrt{4m}$ is a monotonically increasing function of $m$ between 
$g(0) = 0$ and $g(1) = \sqrt{2}$.
This indicates that the solitonic sector for 2DBOI ($m\sim1/2$) 
is more unstable than the periodic sector ($0<m<1/2$), which in turn is more unstable than the linear sector ($m\sim0$).

\smallskip
\paragraph{Numerical validation.}

To check the stability results from the 2DBO-Whitham system, 
we also computed the growth rates for the 2DBOI equation numerically.
Let $u_m(x,y,t)$ 
be a traveling wave solution of the 2DBO equation
as in Eq.~\eqref{e:periodicsoln}, and let
$\xi = x - c t$.
We seek a perturbed solution in the form $u(x,y,t) = u_m(\xi) + v(\xi,y,t)$ with $|v(\xi,y,t)| \ll 1$. 
Substituting this expansion into the 2DBO equation and dropping higher-order terms, we have
$\big(v_t - c v_{\xi} + (u_m v)_{\xi} + \epsilon \mathcal{H} [v_{\xi\xi}]\big)_{\xi} + \lambda v_{yy} = 0$.
Using Galilean invariance, we can transform $u_0$ to $c + \~u_0$, obtaining
(dropping tildes)
\vspace*{-0.6ex}
\begin{align}
\label{e:perturbeqn}
\big(v_t + (u_m v)_{\xi} + \epsilon \mathcal{H} [v_{\xi\xi}]\big)_{\xi} + \lambda v_{yy} = 0\,.
\end{align}
Next we look for plane wave solution of Eq.~\eqref{e:perturbeqn} in the form
$v(\xi,y,t) = w(\xi) \e^{i \zeta y + \mu t}$,
obtaining
$\big(\mu w + (\~u_0 w)_{\xi} + \epsilon \mathcal{H} [w_{\xi\xi}]\big)_{\xi} - \lambda \zeta^2 w = 0$,
or equivalently, 
assuming that $w$ has no mean term,
\vspace*{-1ex}
\begin{align}
\label{e:eigenprob}
- (\~u_0 w)_{\xi} - \epsilon \mathcal{H} [w_{\xi\xi}] + \lambda \zeta^2 \partial^{-1}_{\xi} w = \mu w \,,
\end{align}
where $\partial^{-1}_{\xi} w = \mathcal{F}^{-1} \big[(1/i k)\,\mathcal{F}[w]\big]$ 
and $\mathcal{F}$ denotes the Fourier transform. 
One can now treat Eq.~\eqref{e:eigenprob} as a differential eigenvalue problem, 
which can be efficiently solved numerically in the Fourier domain
using Hill's method~\cite{deconinckkutz}
to obtain the growth rate as the largest imaginary eigenvalue.

\begin{figure}[t!]
\kern-\smallskipamount
\centerline{\includegraphics[width=0.385\textwidth]{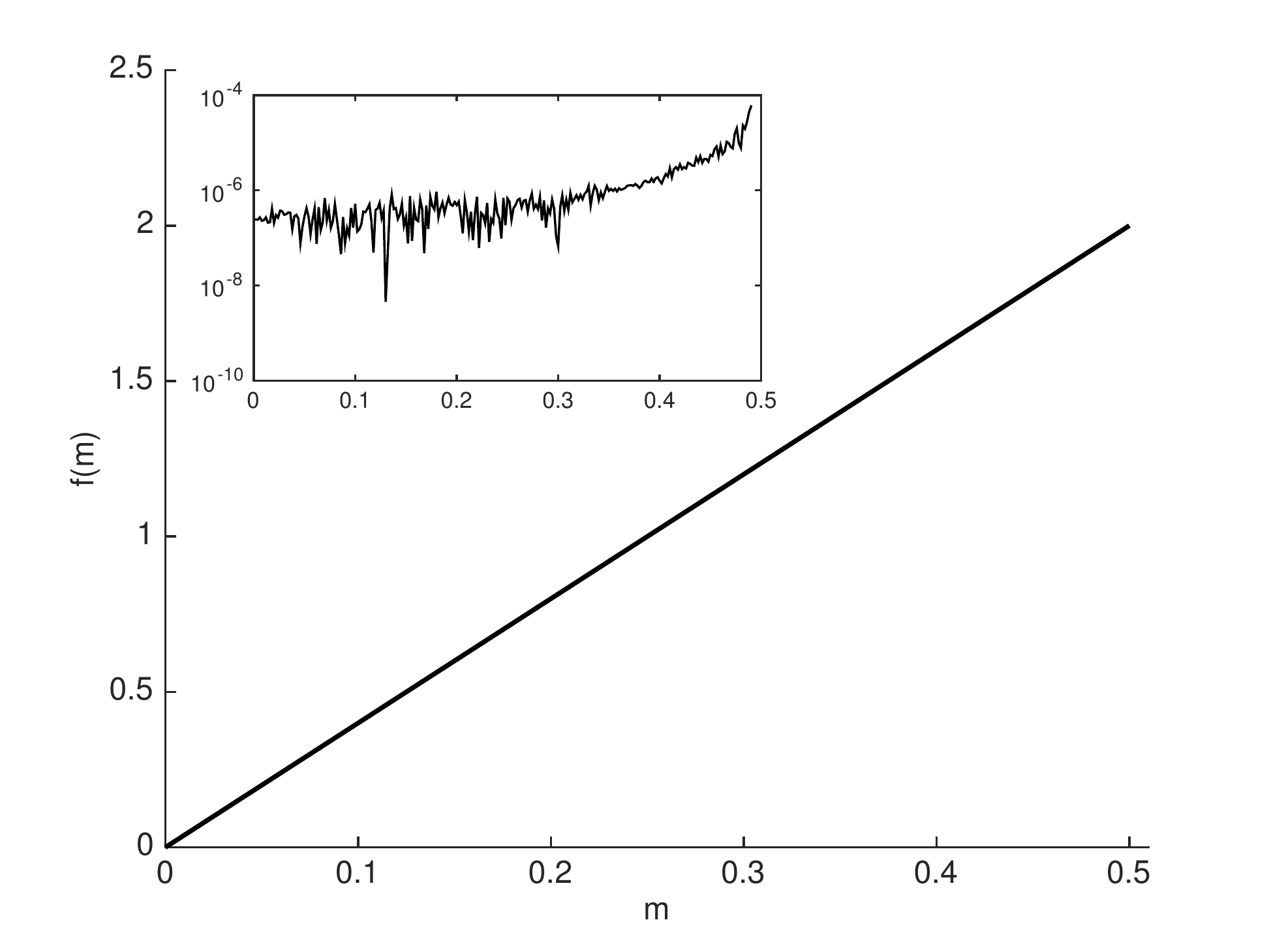}}
\kern-\smallskipamount
\caption{The square $f(m) = 4m$ of the instability growth rate 
as predicted by Whitham modulation theory, given by Eq.~\eqref{e:dispersionrelation2},
Inset: comparison with the numerically computed growth rates.}
\label{f:growth}
\kern-\smallskipamount
\end{figure}

To compare the numerical results with those obtained from Whitham modulation theory, 
note that when $r_1 = q = p = 0$, $r_2 = m$ and $r_3 = 1/2$, the periodic solution~\eqref{e:periodicsoln} becomes
\vspace*{-1ex}
\be
u_m(x,t) = {2a^2}/\{{1-a -  \sqrt{2 m}\cos[a\,(x-(1-a)t)]}\}\,,
\nonumber
\ee
with $a = \frac12 - m$.
We then solved numerically the resulting eigenvalue problem with $0< m <1/2$. 
The difference between the numerically computed growth rates and those obtained from Whitham theory,
shown in Fig.~\ref{f:growth},
is less than $10^{-6}$ for most values of $m$, and less than $10^{-4}$ in all cases,
demonstrating excellent agreement, 
and providing a strong validation of the perturbation expansion presented in section~II 
as well as the usefulness of the 2DBO-Whitham system itself.


\section{V.~ Conclusions}
\label{s:conclusion}

In summary, we studied
the small dispersion limit of the 2DBO equation by deriving a Whitham modulation system. 
We transformed the system to Riemann-type variables and we showed how suitable manipulations
allow one to obtain a system that is free of singularities,
referred to here as the 2DBO-Whitham system.
We discussed several exact reductions of the system  as well as the formulation of well-posed IVPs for the 2DBO-Whitham system, 
including the 2D generalization of the Riemann problem.
We also used the 2DBO-Whitham system to study the linear spectral stability  of the traveling wave solutions of the 2DBO equation and
found that all such solutions are spectrally unstable for the 2DBOI equation and spectrally stable for the 2DBOII equation 
We compared the analytically computed growth rates with a direct numerical approach,  
obtaining excellent agreement.

From a physical point of view,
the above stability results imply that periodic trains of internal waves can be expected to be stable 
to transverse perturbations in stratified media in which surface tension is not dominant
(i.e., media for which the 2DBOII variant of the 2DBO equation is the appropriate model, as opposed to 2DBOI).

The results of this work open up several possibilities for further 
development of the theory.
One such possibility is
whether the methods used in this work can be applied to even further (2+1)-dimensional equations, 
e.g. such as the modified Kadomtsev-Petviashvili equation,
in order to generalize the results obtained in~\cite{driscolloneil} for the modified Korteweg-de\,Vries equation.
On the other hand, regarding the 2DBO equation,
further work is clearly needed to more fully understand the properties of the 2DBO-Whitham system.
Importantly, we also expect that one can
use the 2DBO-Whitham system to study, analytically and numerically, 
the formation of multi-dimensional DSWs in the 2DBO equation.
We plan to address some of these questions in the near future.

\kern\bigskipamount

\subsection*{Acknowledgments}

We are grateful to I. Rumanov for bringing to our attention the alternative derivation of 
some of the modulation equations in the 2DBO-Whitham system, i.e., Eqs.~\eqref{e:pqnew}.
This work was partially supported by the National Science Foundation 
under grant numbers
DMS-1712793, 
DMS-1615524
and 
DMS-1614623.

\def\href#1{\relax}
\def\title#1{``#1''}

\end{document}